\documentclass[twocolumn,amsmath,amssymb,prl,showpacs,floatfix]{revtex4}
\usepackage{graphicx}
\usepackage{dcolumn}
\usepackage{bm}
\begin{document}

\title{Pressure-induced volume-collapsed tetragonal phase of CaFe$_2$As$_2$ \\ as seen via neutron scattering}

\author{A.~Kreyssig$^{1,2}$}
\author{M.~A.~Green$^{3,4}$}
\author{Y.~Lee$^{1,2}$}
\author{G.~D.~Samolyuk$^{1,2}$}
\author{P.~Zajdel$^{3,5}$}
\author{J.~W.~Lynn$^{3}$}
\author{S.~L.~Bud'ko$^{1,2}$}
\author{M.~S.~Torikachvili$^{6}$}
\author{N.~Ni$^{1,2}$}
\author{S.~Nandi$^{1,2}$}
\author{J.~Le\~{a}o$^{3}$}
\author{S.~J.~Poulton$^{3,4}$}
\author{D.~N.~Argyriou$^{7}$}
\author{B.~N.~Harmon$^{1,2}$}
\author{R.~J.~McQueeney$^{1,2}$}
\author{P.~C.~Canfield$^{1,2}$}
\author{A.~I.~Goldman$^{1,2}$}

\affiliation{\\$^1$Ames Laboratory, US DOE; Ames, IA 50011; USA}

\affiliation{$^2$Department of Physics and Astronomy, Iowa State
University; Ames, IA 50011; USA}

\affiliation{$^3$NIST Center for Neutron Research, National 
Institute of Standards and Technology; Gaithersburg, MD 20899; 
USA}

\affiliation{$^4$Department of Materials Science and Engineering, 
University of Maryland; College Park, MD 20742; USA}

\affiliation{$^5$Department of Chemistry, University College of 
London; 20 Gordon Street; London, W1X 0AJ; UK}

\affiliation{$^6$Department of Physics, San Diego State 
University; San Diego, CA 92182; USA}

\affiliation{$^7$Helmholtz-Zentrum Berlin f\"ur Materialien und 
Energie; Glienicker Str. 100; 14109 Berlin; Germany}


\begin{abstract}
Recent investigations of the superconducting iron-arsenide 
families have highlighted the role of pressure, be it chemical or 
mechanical, in fostering superconductivity.  Here we report that 
CaFe$_2$As$_2$ undergoes a pressure-induced transition to a 
non-magnetic, volume ``collapsed'' tetragonal phase, which 
becomes superconducting at lower temperature.  Spin-polarized 
total-energy calculations on the collapsed structure reveal that 
the magnetic Fe moment itself collapses, consistent with the 
absence of magnetic order in neutron diffraction.  
\end{abstract}

\pacs{61.50.Ks, 61.05.fm, 71.15.Nc, 74.62.Fj; Paper accepted for 
publication in Phys. Rev. B}


\maketitle{}

Two recently 
discovered\cite{Kamihara08,Takahashi08,Chen08,Rotter08} series of 
high transition temperature (high-$T_c$) superconductors 
originate from the parent systems $R$FeAsO ($R$~= rare earth) and 
$A$Fe$_2$As$_2$ ($A$~= alkaline earth metal), which are 
tetragonal at room temperature but undergo an orthorhombic 
distortion in the range 100-220~K that is associated with the 
onset of antiferromagnetic 
order\cite{Cruz08,Klauss08,Krellner08,Huang08,Jesche08,Goldman08,Zhao08}.  
Tuning the system via element 
substitution\cite{Takahashi08,Chen08,Rotter08,Chen08a,Sasmal08,Wu08} 
or oxygen deficiency\cite{Ren08,Kito08} suppresses the magnetic 
order and structural distortion in favor of superconductivity 
($T_c$'s up to 55~K), with an overall behavior strikingly similar 
to the high-$T_c$ copper oxide family of superconductors.  

The recent report\cite{Torikachvili08} of pressure-induced 
superconductivity in the parent CaFe$_2$As$_2$ compound opens an 
alternative path to superconductivity.  Pressure suppresses the 
distinct resistivity signature of the high-temperature structural 
and magnetic phase transition from 170~K at ambient 
pressure\cite{Ni08} to 128~K at 0.35~GPa\cite{Torikachvili08}.  
Superconductivity emerges with $T_c$ up to 12~K for pressures 
between 0.23~GPa and 0.86~GPa\cite{Torikachvili08}.  The 
pressure-induced superconductivity in CaFe$_2$As$_2$ was 
confirmed\cite{Park08} and followed by observations of 
superconductivity for BaFe$_2$As$_2$ and SrFe$_2$As$_2$ at 
significantly higher pressures\cite{Alireza08}.  In 
CaFe$_2$As$_2$, a second high-temperature phase transition is 
observed above 0.55~GPa and 104~K by anomalies in the 
resistivity\cite{Torikachvili08}.  However, the nature of the 
phase at temperatures below this transition and its relation to 
the ambient-pressure tetragonal, orthorhombic and 
pressure-induced superconducting phases are as yet unknown. 

Neutron scattering experiments on CaFe$_2$As$_2$ were performed 
to elucidate these issues.  Special attention was paid to 
maintain experimental conditions closest to the reported 
macroscopic measurements and under well-defined hydrostatic 
pressure.  Therefore, the experiments were performed on a 
polycrystalline sample prepared out of approximately 1.75~grams 
of single crystalline CaFe$_2$As$_2$ material grown using the 
procedure described in references \cite{Ni08} and 
\cite{Canfield92}.  The temperature profile for preparing this 
material was slightly modified (heating to 1100$^{\circ}$C and 
cooling over 50~hours to 600$^{\circ}$C) to inhibit the formation 
of the reported\cite{Ni08} needle-shaped impurity phase.  
Temperature-dependent resistance measurements on these crystals 
reproduced the data presented in references \cite{Torikachvili08} 
and \cite{Ni08}.  The single crystals (~300 pieces) were loaded 
with attempted random orientation into a He-gas pressure cell 
(maximum pressure 0.63~GPa) and measured on the BT1 
high-resolution powder diffractometer at the NIST Center for 
Neutron Research.  To reduce the effects of preferred 
orientation, the sample was oscillated over an angle of 36~deg 
during each measurement.  For the temperature-dependent studies, 
the temperature was slowly changed with a maximum rate of 5~K/min 
while the pressure was adjusted and allowed to equilibrate 
between measurements.  

Figure 1 shows neutron diffraction scans taken through the 
nuclear (0~0~2), (2~2~0)$_{\textrm{T}}$, and magnetic 
(1~2~1)$_{\textrm{OR,~magnetic}}$ diffraction peaks at selected 
temperatures and pressures.  At 50~K and ambient pressure (A), 
the splitting of the (2~2~0)$_{\textrm{T}}$ into the orthorhombic 
(4~0~0)$_{\textrm{OR}}$/(0~4~0)$_{\textrm{OR}}$ peaks signals the 
transformation to the orthorhombic phase (Fig.~1(b)). This, 
together with the observation of the magnetic 
(1~2~1)$_{\textrm{OR,~magnetic}}$ peak (Fig.~1(c)), is consistent 
with previous x-ray and neutron diffraction measurements at 
ambient pressure\cite{Goldman08,Ni08}.  Upon increasing pressure 
at $T$~=~50~K, the structure remains orthorhombic and 
antiferromagnetic up to approximately 0.24~GPa.  

Between 0.24 and 0.35~GPa, dramatic changes take place in the 
measured diffraction patterns.  At pressures above 0.35~GPa 
(Fig.~1(c)), the magnetic peak is absent. No other magnetic 
magnetic peaks (e.g. corresponding to the AF1 magnetic phase 
proposed in reference \cite{Yildirim08}) are observed.  The 
orthorhombic structure has transformed to a tetragonal phase, 
similar to the high-temperature ambient pressure structure, but 
with extraordinarily different lattice parameters.  This is most 
evident from the strong shift in the positions of the (0~0~2) and 
(2~2~0)$_{\textrm{T}}$ peaks at (B) in Figs.~1(a) and 1(b), 
respectively.  The structure of this pressure-induced 
``collapsed'' tetragonal phase is unchanged in the 
superconducting state determined by measurements at 4 K and under 
0.48 GPa.  

\begin{figure}
\includegraphics[width=1.0\linewidth]{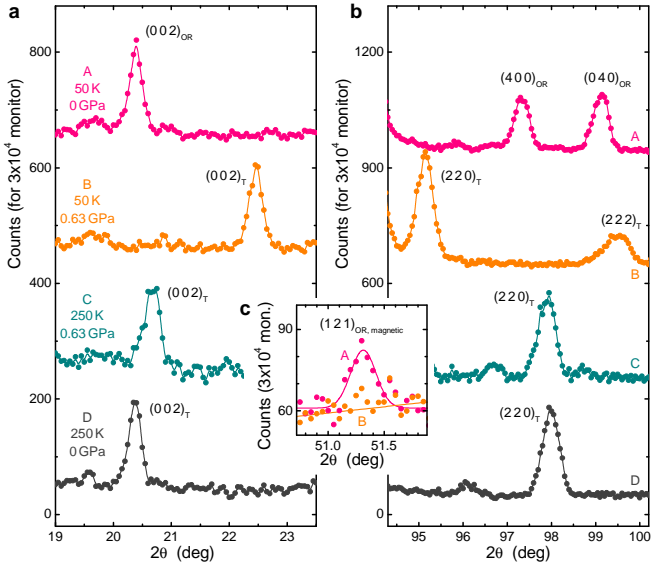}
\caption{\label{fig:fig1} (color online) Scans through (a-b) 
nuclear and (c) magnetic peaks in neutron diffraction pattern at 
selected temperatures and pressures.  Note that the diffraction 
peaks change position dramatically due to the significant changes 
in the lattice parameters.  In (c), the magnetic 
(1~2~1)$_{\textrm{OR,~magnetic}}$ diffraction peak for point A is 
clearly observed above the background taken at B.  No new 
magnetic peaks for B and C were observed.  Unlabelled peaks in 
the pattern arise from phases other than CaFe$_2$As$_2$, such as 
minor contamination from the Sn flux or SiO$_2$ (silica wool or 
pieces of the silica ampule from the single crystal growth), or 
the pressure cell.  The subscripts denote the crystal structure 
used for indexing (T~= tetragonal; OR~= orthorhombic).  The 
offset between every data set is 200~Counts/3x10$^4$ monitor in 
(a) and 300~Counts/3x10$^4$ monitor in (b), respectively.}
\end{figure}

The central region (shown in yellow) of Figs. 2(b) and (c) shows 
the results of Rietveld refinements of the lattice parameters and 
volume for the ``collapsed'' tetragonal phase.  The structure 
data (lattice parameters and atomic-position parameter 
$z_{\textrm{As}}$ of As) were determined by Rietveld refinements 
using the GSAS software package\cite{Larson94}.  We find an 
astonishing 9.5\% reduction in the $c$-lattice parameter with 
respect to the orthorhombic phase and a nearly 5\% decrease in 
the unit cell volume.  Even more striking is the reduction of the 
$c/a$ ratio, a key parameter for bond geometries in the iron 
arsenides, by nearly 11\%.  As a consequence, the As-Fe-As bond 
angles change strongly as illustrated in Fig.~2(d).  

\begin{figure}
\includegraphics[width=1.0\linewidth]{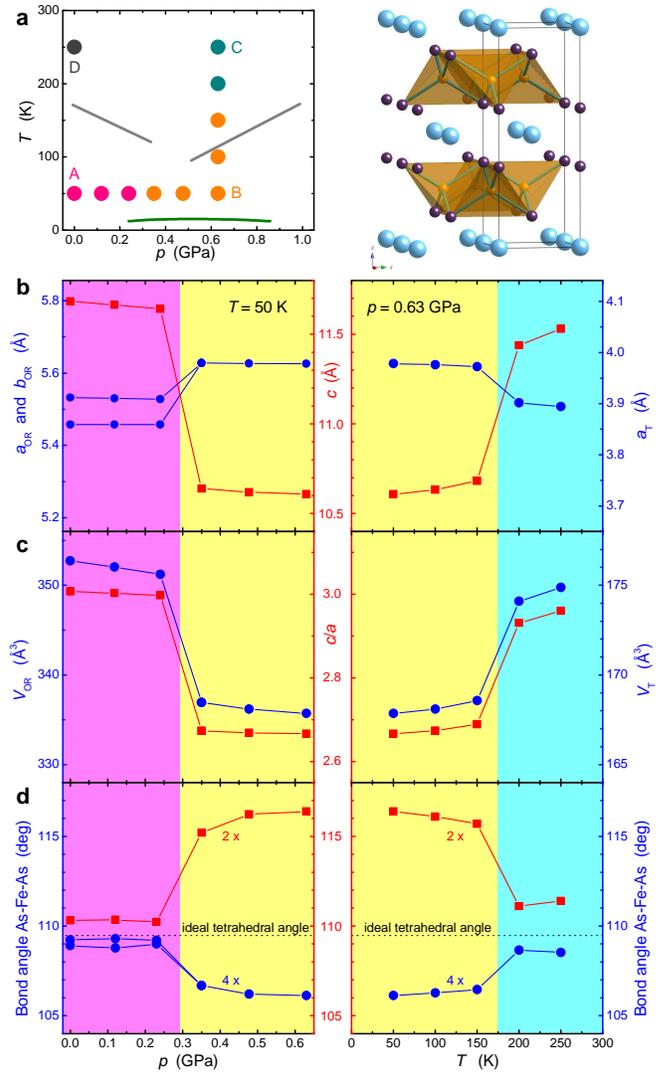}
\caption{\label{fig:fig2} (color online) Schematic phase diagram 
and pressure and temperature dependence of the lattice 
parameters, unit cell volume and As-Fe-As bond angles. (a) (left 
panel) Lines in the schematic $p-T$ diagram denote the 
high-temperature and superconductivity phase lines determined in 
reference \cite{Torikachvili08}.  Points A-D label the pressures 
and temperatures for the diffraction data shown in Figure 1.  (a) 
(right panel) The unit cell of the tetragonal phase of 
CaFe$_2$As$_2$.  (b-d) Pressure dependence (left panels) at 
$T$~=~50~K and temperature dependence (right panels) at 
$p$~=~0.63~GPa (b) of the lattice parameters, (c) of the unit 
cell volume and $c/a$ ratio, and (d) of the As-Fe-As bond angle.}
\end{figure}

With the pressure maintained at 0.63~GPa, the temperature was 
raised in 50~K steps (right panels of Fig.~2).  Between 150 and 
200~K an isostructural transition between the low-temperature 
``collapsed'' tetragonal phase and the high-temperature 
tetragonal structure is observed.  Upon release of the pressure 
at 250~K, the curves labelled (D) in Fig.~1 show only small 
changes in the lattice parameters between 0.63~GPa and ambient 
pressure, providing a measure of the modest, but strongly 
anisotropic compressibility of the high-temperature phase.  

We note that there is a difference of about 50~K between the 
temperature of the isostructural transition at 0.63~GPa measured 
here and that reported in transport 
measurements\cite{Torikachvili08}.  However, as pointed out in 
Ref. \cite{Torikachvili08}, the resistive anomalies are rather 
broad in applied pressure, and different criteria for the 
definition of transition temperatures can shift temperature 
assignments. In addition, the data in Ref. \cite{Torikachvili08} 
were taken with decreasing temperature whereas here the 
temperature was stepwise increased.  With these uncertainties 
understood, the tetragonal-to-``collapsed'' tetragonal transition 
appears to be responsible for the loss of resistivity whose locus 
defines the high-temperature high-pressure phase line found in 
Ref. \cite{Torikachvili08} and shown in Fig.~2(a).

In order to relate the volume change to relative changes in the 
unit cell dimensions, and to verify the stability of this phase, 
spin-polarized total-energy calculations were performed for 
volume changes of $\Delta V/V = 0\%$ (for ambient pressure) and 
$\Delta V/V = -5\%$ (for the ``collapsed'' phase).  The local 
density approximation was employed, using the full potential 
linearized augmented plane wave method with the Perdew-Wang 1992 
functional\cite{Perdew92}.  The precision of the total energy is 
0.01~mRyd/cell, much smaller than the size of the symbols in 
Fig.~3.  

From the blue (dark grey) curves in Fig.~3(a) we see that, for 
ambient pressure, the orthorhombic magnetic phase is lowest in 
energy, consistent with our ambient-pressure low-temperature 
measurement.  The red (light grey) curves in Fig.~3(a) show that 
the tetragonal phase is lowest in total energy for the 5\% volume 
reduction.  The minimum energy of this ``collapsed'' tetragonal 
phase is found at $c/a~\sim~2.65$, close to the experimental 
value of 2.67 (Fig.~2(c)).  

The $c/a$-dependence of the spin-polarized total-energy 
calculations for the ``collapsed'' phase can be correlated with a 
loss of the Fe magnetic moment, as shown in Fig.~3(b).  Both the 
spin-polarized and non-spin-polarized calculations yield the same 
total-energy minimum for the non-magnetic ``collapsed'' 
tetragonal phase at the same $c/a$ ratio.  The astonishing result 
of a quenched magnetic moment ground state is consistent with our 
experimental observation of the loss of magnetic order in the 
``collapsed'' tetragonal phase.  The band structure calculations 
also indicate that several bands cross the Fermi level at the 
pressure-induced transition.  

\begin{figure}
\includegraphics[width=1.0\linewidth]{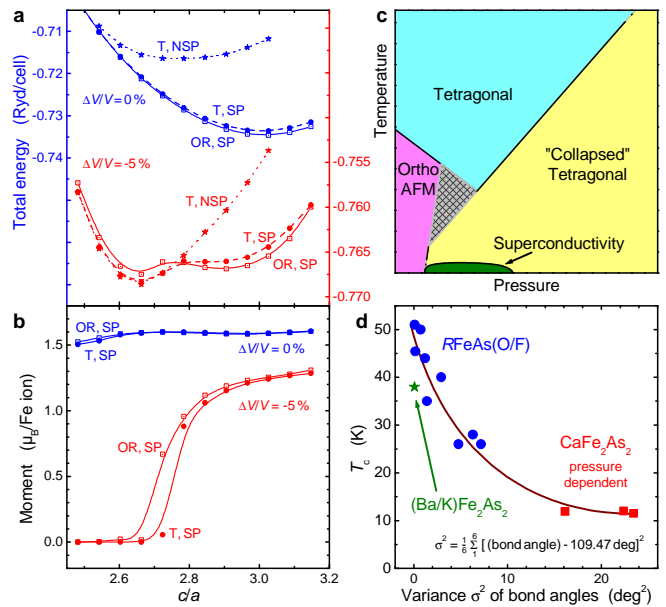}
\caption{\label{fig:fig3} (color online) Summary of the results 
of the neutron diffraction measurements and total-energy 
calculations. (a) Spin-polarized (SP) and non-spin-polarized 
(NSP) total-energy calculations for $\Delta V/V = 0\%$ and 
$\Delta V/V = -5\%$ for the tetragonal (T) and orthorhombic (OR) 
phases (Note the different energy scales).  The tetragonal 
notation is used for the $c/a$ ratio.  (b) For the ``collapsed'' 
phase, the Fe moment is quenched at the minimum in total energy 
in the spin-polarized calculation.  (c) Schematic $p-T$ diagram 
based upon the diffraction and transport\cite{Torikachvili08} 
measurements.  Dashed lines represent estimates of the phase 
boundaries.  The cross hatched area indicates a region where 
precise details of how these transition lines intersect still 
needs to be determined.  (d) Correlation between the 
superconducting transition temperatures and the variance 
$\sigma^2$ of the As-Fe-As bond angle.  The data were obtained 
from Refs. \cite{Cruz08} and 
\cite{Zhao08a,Qiu08,Qiu08a,Bos08,Lee08} for the $R$FeAs(O/F) 
compounds and from Ref. \cite{Rotter08} for (Ba/K)Fe$_2$As$_2$, 
respectively.  The data for CaFe$_2$As$_2$ are derived from the 
present measurements.}
\end{figure}

The principal result of these neutron diffraction measurements is 
the discovery of a transition from the magnetically ordered 
orthorhombic phase to a non-magnetically ordered ``collapsed'' 
tetragonal phase preceding the onset of superconductivity.  
Further, the second, higher pressure, transition noted in 
transport measurements\cite{Torikachvili08} has been identified 
as an isostructural transition between the pressure-induced 
``collapsed'' phase and the high-temperature tetragonal 
structure.  The observed volume reduction can, for example, serve 
to increase the charge-carrier density.  The schematic phase 
diagram in Fig.~3(c) summarizes our findings.  Our results show 
that the pressure-induced superconductivity\cite{Torikachvili08} 
in CaFe$_2$As$_2$ emerges from the ``collapsed'' tetragonal phase 
rather than the magnetically ordered, orthorhombic phase or the 
high-temperature tetragonal phase. 

Anomalous changes in the unit cell volume and lattice constants 
have also been noted in the $R$FeAsO system.  In superconducting 
fluorine-free oxygen-deficient samples of NdFeAsO$_{1-\delta}$, a 
surprising discontinuous decrease in the lattice parameters and 
unit cell volume ($\Delta V/V = -1.8\%$) was found for 
$\delta$~=~0.4, where a maximum in the superconducting volume 
fraction is observed\cite{Kito08}.  Furthermore, a correlation 
between superconducting transition temperatures for 
$R$FeAsO$_{1-\delta}$ and the unit cell dimensions is 
reported\cite{Ren08}.  In light of our results, it appears that 
chemical substitution and the introduction of oxygen deficiency 
likely play a dual role in the iron arsenide superconductors by 
increasing the charge carrier density and changing the ``chemical 
pressure''.  It is not yet clear which has the greater impact 
upon superconductivity.  

Given that the $R$FeAsO and $A$Fe$_2$As$_2$ families share a 
common structural element (FeAs layers) and similar prerequisites 
for superconductivity (e.g. suppression of magnetic order) it is 
useful to elucidate structural quantities that are shared and can 
be correlated with superconducting properties.  Of particular 
interest is the comparison of the As-Fe-As bond 
angles\cite{Lee08} with the ideal tetrahedral value of 
109.47~deg.  For this special value all Fe atoms are coordinated 
in ideally formed tetrahedrons with identical high-symmetric 
Fe-As-Fe bonding geometries [highlighted in brown in the right 
panel of Fig.~2(a)].  This symmetry is broken for As-Fe-As bond 
angles deviating from this ideal angle yielding two or three 
different values in the tetragonal or orthorhombic structure, 
respectively.  The variance $\sigma^2=\frac{1}{6}\sum \limits 
_1^6 [(\textrm{bond~angle})-109.47~\textrm{deg}]^2$ in deviation 
of the Fe-As-Fe bonding angles from the ideal value parameterizes 
the strength of this symmetry breaking.  

In Figure 3(d) we plot the measured superconducting transition 
temperatures $T_c$, for those iron-arsenide compounds that are 
superconducting, as a function of this variance $\sigma^2$ of the 
As-Fe-As bond angles.  As $\sigma^2$ increases, corresponding to 
greater deviations in the bond angle from the ideal tetrahedral 
angle, $T_c$ decreases.  The pressure-dependent properties of the 
``collapsed'' tetragonal phase of CaFe$_2$As$_2$ described above 
clearly continue this trend and allow generalizing our result to 
the family of FeAs based superconductors.  The high value for 
$\sigma^2$ in the ``collapsed'' tetragonal phase for 
CaFe$_2$As$_2$ is consistent with the low $T_c$ in comparison to 
other FeAs based superconductors.  The observed correlation 
between $T_c$ and $\sigma^2$ points to the importance of the 
structure and the symmetry in the FeAs network for the 
superconducting state.  Properties that are sensitive to the 
As-Fe-As bonding geometry and its symmetry, such as anisotropic 
magnetic or elastic couplings in the FeAs network, seem strongly 
involved in the superconducting pairing.  

These results highlight intriguing questions that point to the 
potential complexity of the superconducting state in the iron 
arsenides.  From the analysis of $T_c$ as a function of the 
As-Fe-As bond angle variance, we have found that 
superconductivity in the ``collapsed'' tetragonal phase fits well 
within the general trend observed for the doped iron arsenides, 
implying a common superconducting pairing mechanism.  It has been 
suggested that spin fluctuations are responsible for the electron 
pairing in this class of superconductors\cite{Dong08}.  The 
apparent loss of a static moment in the ``collapsed'' tetragonal 
phase may seem inconsistent with such a magnetic pairing 
mechanism for the pressure-induced superconductivity in 
CaFe$_2$As$_2$.  However, the strong pair-breaking effect of 
local moments is eliminated and superconductivity mediated by 
paramagnons\cite{Chubukov02} remains a possibility. 
\\

The authors wish to acknowledge very useful discussions with 
Joerg Schmalian.  Work at the Ames Laboratory was supported by 
the US Department of Energy - Basic Energy Sciences under 
Contract No. DE-AC02-07CH11358.  M.~S.~Torikachvili acknowledges 
support by the National Science Foundation under DMR-0306165 and 
DMR-0805335. 

\section{Appendix A: Details of the neutron diffraction measurements and their 
analysis including structure data} 

The neutron diffraction measurements were performed on the high 
resolution powder diffractometer BT1 at the NIST Center for 
Neutron Research using a wavelength of 2.0782~$\AA$ selected by a 
Ge (3~1~1) monochromator.  The collimation of the incident beam 
was set at 15'.

Special attention was paid to maintaining experimental conditions 
closest to the reported macroscopic measurements and to perform 
the study under well-defined hydrostatic pressure.  The soft and 
ductile nature of the compound presents challenges for powder 
diffraction measurements since flux-grown single crystals tend to 
smear and shear when ground into a powder.  The effects of 
grinding are clearly observed as broadened peaks in x-ray powder 
diffraction measurements, and the modification of physical 
properties associated with grinding have not yet been 
characterized.  Therefore, as grown single crystals 
($\sim$500-1000) were loaded into an Al-alloy He-gas pressure 
cell and cooled using a top-loading closed-cycle cryogenic 
refrigerator.  The pressure cell was connected to a pressurizing 
intensifier through a high pressure capillary.  Hydrostatic 
pressure was maintained throughout the measurements since the 
temperature was kept well above the melting curve for helium (for 
pressures up to 0.63~GPa).  For the temperature dependence 
studies, the temperature was slowly changed with a maximum rate 
of 5~K/min while the pressure was adjusted and allowed to 
equilibrate between measurements.

To reduce the effects of preferred orientation, the sample was 
oscillated over an angle of 36~deg during each measurement.  
Nevertheless, a degree of preferred orientation remained but was 
adequately modelled in subsequent Rietveld refinements using the 
GSAS software package.  For each diffraction pattern 
approximately 70 reflections were used to refine the lattice 
parameters, the z coordinate of the As ions and the 12 parameters 
associated with corrections for preferred orientation.  We point 
out that the correction for preferred orientation remained 
constant over all pressures and temperatures measured and were 
taken as constants in the fits.  Typical R-values (wRp), 
representing the goodness-of-fit, were between 4-5\% for all fits 
demonstrating the accuracy of the model employed.

\begin{table*}
\centering \caption{Results of Rietveld refinements for 
pressure-dependent measurements at a temperature $T$~=~50~K} 
\label{tab:Rietveld50K}
\begin{ruledtabular}
\begin{tabular}{c|cccccc}Pressure $p$ (GPa)& 0 & 0.115 & 0.24 & 0.35 & 0.47 & 0.63\\
\hline
$a (\AA)$ & 5.5312(2) & 5.5294(2) & 5.5275(2) & 3.9792(1) & 3.9785(1) & 3.9780(1)\\
$b (\AA)$ & 5.4576(2) & 5.4577(2) & 5.4575(2)\\
$c (\AA)$ & 11.683(1) & 11.6625(8) & 11.6391(9) & 10.6379(6) & 
10.6178(7) & 10.6073(7)\\
Volume $V (\AA^3)$ & 352.68(4) & 351.94(3) & 351.11(3) & 
168.44(1) & 168.07(1) & 167.85(1)\\
$z_{\textrm{As}}$ & 0.3689(5) & 0.3693(4) & 0.3690(4) & 0.3687(7) 
& 0.3657(7) & 0.3663(5)\\
Fe-As $(\AA)$ & 2.388(6) & 2.3891(3) & 2.3854(3) & 2.3560(4) & 
2.343(3) & 2.340(3)\\
Fe-Fe $(\AA)$ (2x) & 2.7656(1) & 2.76469(8) & 2.7637(1)\\
Fe-Fe $(\AA)$ (2x) & 2.7288(1) & 2.72882(7) & 2.72874(8)\\
Fe-Fe $(\AA)$ (4x) & & & & 2.8137(9) & 2.8132(1) & 2.8128(1)\\
\end{tabular}
\end{ruledtabular}
\end{table*}

\begin{table*}
\centering \caption{Results of Rietveld refinements for 
pressure-dependent measurements at a pressure $p$~=~0.63~GPa} 
\label{tab:Rietveld063GPa}
\begin{ruledtabular}
\begin{tabular}{c|ccccccc}Temperature $T$ (K) & 4 ($p$ = 0.47 GPa) & 50 & 100 & 150 & 200 & 250 & 250 ($p$ = 0 GPa)\\
\hline
$a (\AA)$ & 3.9797(3) & 3.9780(1) & 3.9760(3) & 3.9724(2) & 3.9015(2) & 3.8944(2) & 3.8915(1)\\
$c (\AA)$ & 10.628(1) & 10.6073(7) & 10.633(1) & 10.683(1) & 11.438(2) & 11.530(2) & 11.690(1)\\
Volume $V (\AA^3)$ & 168.33(3) & 167.85(1) & 167.95(3) & 168.57(3) & 174.11(3) & 174.87(3) & 177.04(2)\\
$z_{\textrm{As}}$ & 0.365(2) & 0.3663(5) & 0.3668(9) & 0.3669(9) 
& 0.367(1) & 0.3652(8) & 0.372(1) \\
Fe-As $(\AA)$ & 2.333(9) & 2.340(3) & 2.344(5) & 2.346(5) & 2.365(9) & 2.357(5) & 2.410(9)\\
Fe-Fe $(\AA)$ (4x) & 2.8141(2) & 2.8128(1) & 2.8115(2) & 2.8089(2) & 2.7588(2) & 2.7538(1) & 2.7517(1)\\
\end{tabular}
\end{ruledtabular}
\end{table*}

\begin{figure}
\includegraphics[width=1.0\linewidth]{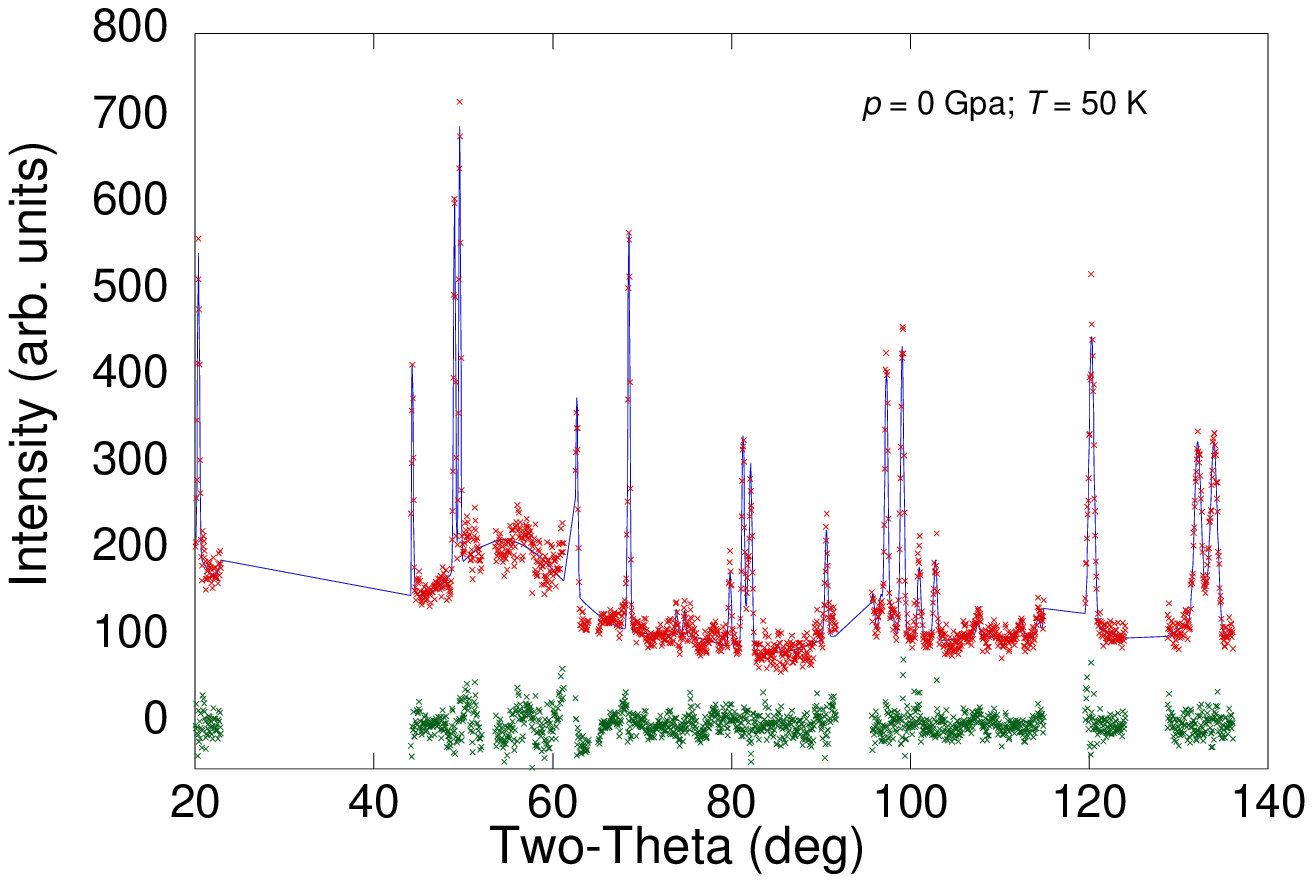}
\caption{\label{fig:fig4} (color online) Rietveld analysis of the 
neutron diffraction pattern at $p$~=~0~Gpa and $T$~=~50~K. Areas 
with strong contributions from the pressure cell are excluded.}
\end{figure}

\begin{figure}
\includegraphics[width=1.0\linewidth]{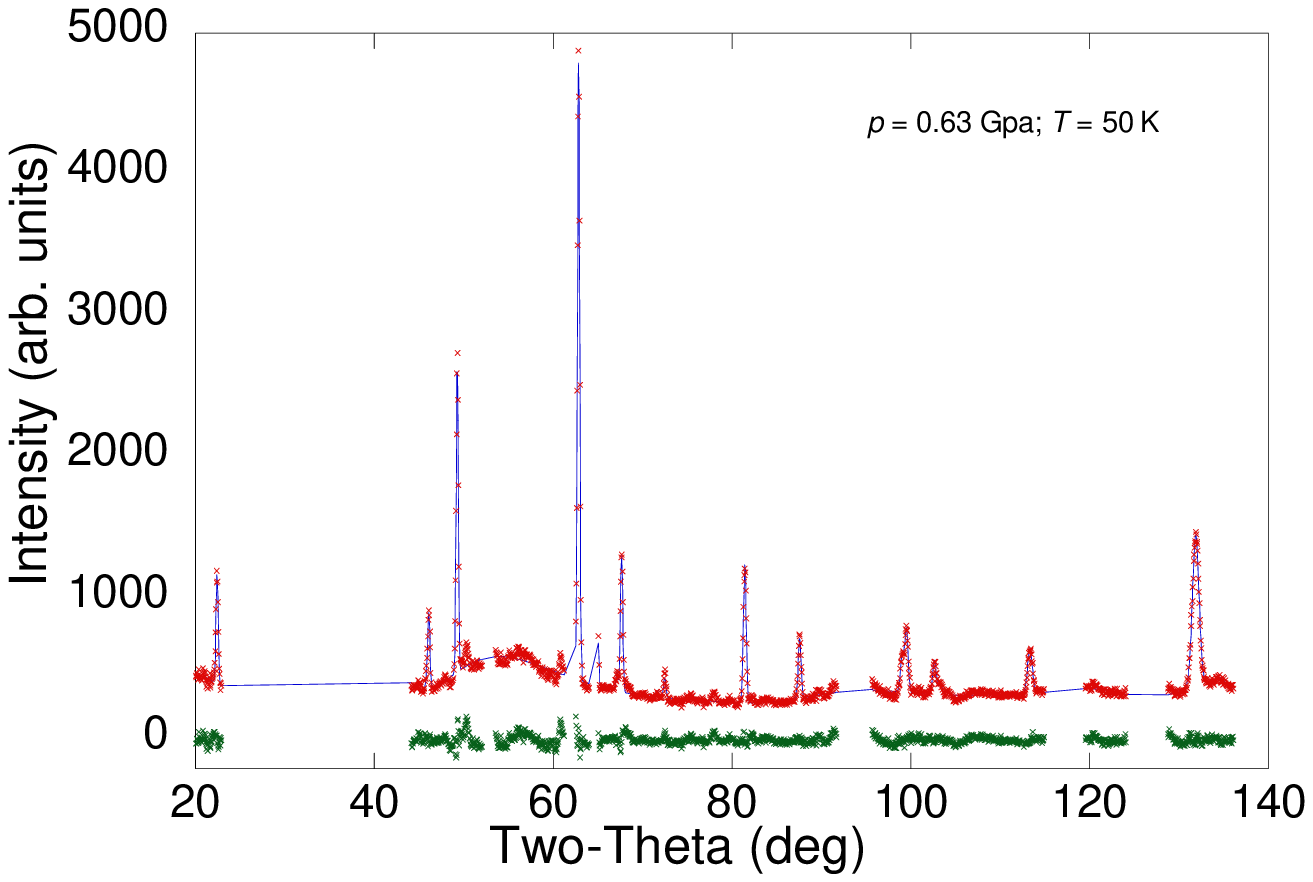}
\caption{\label{fig:fig5} (color online) Rietveld analysis of the 
neutron diffraction pattern at $p$~=~0.63~Gpa and $T$~=~50~K. 
Areas with strong contributions from the pressure cell are 
excluded.}
\end{figure}

\begin{figure}
\includegraphics[width=1.0\linewidth]{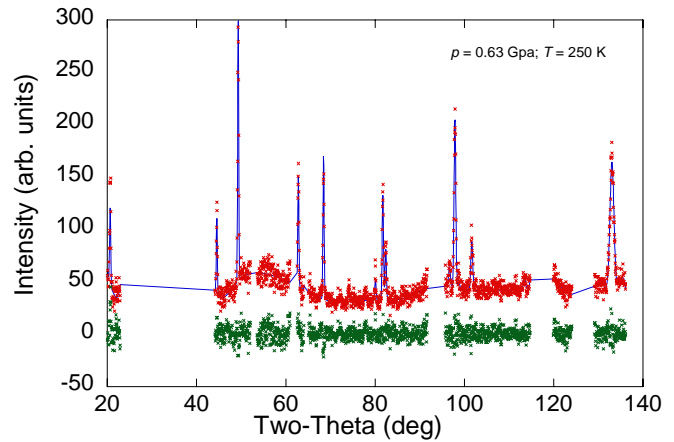}
\caption{\label{fig:fig6} (color online) Rietveld analysis of the 
neutron diffraction pattern at $p$~=~0.63~Gpa and $T$~=~250~K. 
Areas with strong contributions from the pressure cell are 
excluded.}
\end{figure}

\section{Appendix B: Details of the spin-polarized total energy calculations} 

For the spin-polarized and non-spin-polarized calculations, the 
local density approximation was employed, using the full 
potential linearized augmented plane wave method with the 
Perdew-Wang 1992 functional.  The convergence criterion for the 
total energy was 0.01~mRyd/cell.  The calculations were done in 
the tetragonal phase(two formula units) for two different cell 
volumes as the $c/a$ ratio was varied: 0\% volume reduction using 
the experimentally determined lattice parameters and for a 5\% 
volume reduction.  The calculations were also performed for the 
orthorhombic phase, with $a/b$~=~1.02, for both 0\% and 5\% 
volume reduction.  The experimental value of the internal 
parameter was used for the calculations.  The 
$R_{\textrm{MT}}$*$K_{\textrm{max}}$ that determines matrix size 
(the number of the basis functions), where $R_{\textrm{MT}}$ is 
the smallest of all atomic sphere radii and $K_{\textrm{max}}$ is 
the plane wave cut-off was set to 8.0.  The muffin-tin radii 
($R_{\textrm{MT}}$) for the 0\% volume reduction calculation were 
2.2, 2.1 and 2.1 atomic unit for Ca, Fe and As, respectively.  
For the 5\% reduction calculation, the $R_{\textrm{MT}}$ also 
were scaled to keep the matrix size the same.  There were 100 $k$ 
points used in the irreducible Brillouin zone.


\begin{thebibliography}{}

\bibitem[()]{Kamihara08}Y.~Kamihara, T.~Watanabe, M.~Hirano, and H.~Hosono, J. Am. Chem. Soc. \textbf{130}, 3296 (2008). 
\bibitem[()]{Takahashi08}H.~Takahashi, K.~Igawa, K.~Arii, Y.~Kamihara, M.~Hirano, and H.~Hosono, Nature (London) \textbf{453}, 376 (2008).
\bibitem[()]{Chen08}X.~H.~Chen, T.~Wu, G.~Wu, R.~H.~Liu, H.~Chen, and D.~F.~Fang, Nature (London) \textbf{453}, 761 (2008).
\bibitem[()]{Rotter08}M.~Rotter, M.~Tegel, and D.~Johrendt, Phys. Rev. Lett. \textbf{101}, 107006 (2008).
\bibitem[()]{Cruz08}C.~de~la~Cruz, Q.~Huang, J.~W.~Lynn, J.~Li, W.~Ratcliff~II, J.~L.~Zarestky, H.~A.~Mook, G.~F.~Chen, J.~L.~Luo, N.~L.~Wang, and P.~Dai, Nature (London) \textbf{453}, 899 (2008).
\bibitem[()]{Klauss08}H.-H.~Klauss, H.~Luetkens, R.~Klingeler, C.~Hess, F.~J.~Litterst, M.~Kraken, M.~M.~Korshunov, I.~Eremin, S.-L.~Drechsler, R.~Khasanov, A.~Amato, J.~Hamann-Borrero, N.~Leps, A.~Kondrat, G.~Behr, J.~Werner, and B.~B\"uchner, Phys. Rev. Lett. \textbf{101}, 077005 (2008).
\bibitem[()]{Krellner08}C.~Krellner, N.~Caroca-Canales, A.~Jesche, H.~Rosner, A.~Ormeci, and C.~Geibel, Phys. Rev. B \textbf{78}, 100504(R) (2008).
\bibitem[()]{Huang08}Q.~Huang, Y.~Qiu, W.~Bao, M.~A.~Green, J.~W.~Lynn, and Y.~C.~Gasparovic, T.~Wu, G.~Wu, X.~H.~Chen, arXiv:0806.2776 (2008) (unpublished).
\bibitem[()]{Jesche08}A.~Jesche, N.~Caroca-Canales, H.~Rosner, H.~Borrmann, A.~Ormeci, D.~Kasinathan, K.~Kaneko, H.~H.~Klauss, H.~Luetkens, R.~Khasanov, A.~Amato, A.~Hoser, C.~Krellner, and C.~Geibel, arXiv:0807.0632 (2008) (unpublished).
\bibitem[()]{Goldman08}A.~I.~Goldman, D.~N.~Argyriou, B.~Ouladdiaf, T.~Chatterji, A.~Kreyssig, S.~Nandi, N.~Ni, S.~L.~Bud'ko, P.~C.~Canfield, and R.~J.~McQueeney, Phys. Rev. B \textbf{78}, 100506(R) (2008).
\bibitem[()]{Zhao08}J.~Zhao, W.~Ratcliff~II, J.~W.~Lynn, G.~F.~Chen, J.~L.~Luo, N.~L.~Wang, J.~Hu, and P.~Dai, Phys. Rev. B \textbf{78}, 140504(R) (2008).
\bibitem[()]{Chen08a}G.-F.~Chen, Z.~Li, G.~Li, W.-Z.~Hu, J.~Dong, J.~Zhou, X.-D.~Zhang, P.~Zheng, N.-L.~Wang, and J.-L.~Luo, Chin. Phys. Lett. \textbf{25}, 3403 (2008).
\bibitem[()]{Sasmal08}K.~Sasmal, B.~Lv, B.~Lorenz, A.~M.~Guloy, F.~Chen, Y.-Y.~Xue, and C.-W.~Chu, Phys. Rev. Lett. \textbf{101}, 107007 (2008).
\bibitem[()]{Wu08}G.~Wu, H.~Chen, T.~Wu, Y.~L.~Xie, Y.~J.~Yan, R.~H.~Liu, X.~F.~Wang, J.~J.~Ying, and X.~H.~Chen, J. Phys.: Condens. Matter \textbf{20}, 422201 (2008).
\bibitem[()]{Ren08}Z.-A.~Ren, G.-C.~Che, X.-L.~Dong, J.~Yang, W.~Lu, W.~Yi, X.-L.~Shen, Z.-C.~Li, L.-L.~Sun, F.~Zhou, and Z.-X.~Zhao, Europhys. Lett. \textbf{83}, 17002 (2008).
\bibitem[()]{Kito08}H.~Kito, H.~Eisaki, and A.~Iyo, J. Phys. Soc. Jpn. \textbf{77}, 063707 (2008).
\bibitem[()]{Torikachvili08}M.~S.~Torikachvili, S.~L.~Bud'ko, N.~Ni, and P.~C.~Canfield, Phys. Rev. Lett. \textbf{101}, 057006 (2008).
\bibitem[()]{Ni08}N.~Ni, S.~Nandi, A.~Kreyssig, A.~I.~Goldman, E.~D.~Mun, S.~L.~Bud'ko, and P.~C.~Canfield, Phys. Rev. B \textbf{78}, 014523 (2008).
\bibitem[()]{Park08}T.~Park, E.~Park, H.~Lee, T.~Klimczuk, E.~D.~Bauer, F.~Ronning, and J.~D.~Thompson, J. Phys.: Condens. Matter \textbf{20}, 322204 (2008).
\bibitem[()]{Alireza08}P.~L.~Alireza, J.~Gillett, Y.~T.~C.~Ko, S.~E.~Sebastian, and G.~G.~Lonzarich, arXiv:0807.1896 (2008) (unpublished).
\bibitem[()]{Canfield92}P.~C.~Canfield and Z.~Fisk, Phil. Mag. B \textbf{65}, 1117 (1992). 
\bibitem[()]{Yildirim08}T.~Yildirim, arXiv:0807.3936 (2008) (unpublished).
\bibitem[()]{Larson94}A.~C.~Larson and R.~B.~von~Dreele, GSAS (General Structure Analysis System). Los Alamos National Laboratory Report LA-UR-86-748 (1994). 
\bibitem[()]{Perdew92}J.~P.~Perdew and Y.~Wang, Phys. Rev. B \textbf{45}, 13244 (1992). 
\bibitem[()]{Lee08}C.~H.~Lee, A.~Iyo, H.~Eisaki, H.~Kito, M.~T.~Fernandez-Diaz, T.~Ito, K.~Kihou, H.~Matsuhata, M.~Braden, and K.~Yamada, J. Phys. Soc. Jpn. \textbf{77}, 083704 (2008).
\bibitem[()]{Zhao08a}J.~Zhao, Q.~Huang, C.~de~la~Cruz, S.~Li, J.~W.~Lynn, Y.~Chen, M.~A.~Green, G.~F.~Chen, G.~Li, Z.~Li, J.~L.~Luo, N.~L.~Wang, and P.~Dai, arXiv:0806.2528 (2008) (unpublished). 
\bibitem[()]{Qiu08}Y.~Qiu, W.~Bao, Q.~Huang, T.~Yildirim, J.~M.~Simmons, M.~A.~Green, J.~W.~Lynn, Y.~C.~Gasparovic, J.~Li, T.~Wu, G.~Wu, and X.~H.~Chen, arXiv:0806.2195 (2008) (unpublished). 
\bibitem[()]{Qiu08a}Y.~Qiu, M.~Kofu, W.~Bao, S.-H.~Lee, Q.~Huang, T.~Yildirim, J.~R.~D.~Copley, J.~W.~Lynn, T.~Wu, G.~Wu, and X.~H.~Chen, Phys. Rev. B \textbf{78}, 052508 (2008). 
\bibitem[()]{Bos08}J.-W.~G.~Bos, G.~B.~S.~Penny, J.~A.~Rodgers, D.~A.~Sokolov, A.~D.~Huxleyac, and J.~P.~Attfield, Chem. Commun., 3634 (2008). 
\bibitem[()]{Dong08}J.~Dong, H.~J.~Zhang, G.~Xu, Z.~Li, G.~Li, W.~Z.~Hu, D.~Wu, G.~F.~Chen, X.~Dai, J.~L.~Luo, Z.~Fang, and N.~L.~Wang, Europhys. Lett. \textbf{83}, 27006 (2008).
\bibitem[()]{Chubukov02}A.~Chubukov, D.~Pines, and J.~Schmalian, In: K.-H.~Bennemann and J.~B.~Ketterson (eds.), The Physics of Superconductors. Springer (2002).

\end{thebibliography}
\end{document}